\documentclass[graybox]{svmult}

\usepackage{helvet}         
\usepackage{courier}        
\usepackage{type1cm}        
\usepackage{makeidx}         
\usepackage{graphicx}
\usepackage[bottom]{footmisc}
\usepackage{bm}
\usepackage{amssymb}
\usepackage{float}
\usepackage{amsmath}
\usepackage{dcolumn}

\newcommand{\HCd}{\mathcal{H}}

\def\HCdt0{\tilde{\HCd}_{0}}

\usepackage{tensor}
\usepackage{mathbbol}
\newcommand\rf[1]{(\ref{eq:#1})}
\newcommand\lab[1]{\label{eq:#1}}
\newcommand\nonu{\nonumber}
\newcommand\br{\begin{eqnarray}}
\newcommand\er{\end{eqnarray}}
\newcommand\be{\begin{equation}}
\newcommand\ee{\end{equation}}

\newcommand\lb{\lbrack}
\newcommand\rb{\rbrack}

\renewcommand\({\left(}
\renewcommand\){\right)}

\newcommand\bc{\begin{center}}
\newcommand\ec{\end{center}}

\renewcommand\a{\alpha}

\renewcommand\d{\delta}

\newcommand\vareps{\varepsilon}

\newcommand\G{\Gamma}

\newcommand\h{\frac{1}{2}}
\renewcommand\k{\kappa}
\renewcommand\l{\lambda}
\renewcommand\L{\Lambda}
\newcommand\m{\mu}
\newcommand\n{\nu}
\newcommand\om{\omega}

\newcommand\vp{\varphi}
\renewcommand\P{\Phi}
\newcommand\pa{\partial}

\newcommand\pr{\prime}

\newcommand\s{\sigma}

\newcommand\cA{{\mathcal A}}
\newcommand\cB{{\mathcal B}}

\newcommand\cL{{\mathcal L}}
\newcommand\cM{{\mathcal M}}

\newcommand\cT{{\mathcal T}}

\newcommand{\ct}[1]{\cite{#1}}
\newcommand\PRL[3]{{Phys. Rev. Lett.} \textbf{#1}, #3 (#2)}

\newcommand\PRD[3]{{Phys. Rev.} \textbf{D#1}, #3 (#2)}

\newcommand\AoP[3]{{Ann. of Phys.} \textbf{#1}, #3 (#2)}

\newcommand\PRep[3]{{Phys. Reports} \textbf{#1}, #3 (#2)}

\newcommand\IJMPA[3]{{Int. J. Mod. Phys.} \textbf{A#1}, #3 (#2)}
\newcommand\IJMPD[3]{{Int. J. Mod. Phys.} \textbf{D#1}, #3 (#2)}

\newcommand\MPLA[3]{{Mod. Phys. Lett.} \textbf{A#1}, #3 (#2)}

\newcommand\vpdot{\stackrel{.}{\varphi}}

\newcommand\addot{\stackrel{..}{a}}

\begin{document}

\title*{Modified Gravity Theories Based on the Non-Canonical Volume-Form Formalism}
\titlerunning{Modified Gravity Theories from Non-Canonical Volume-Forms}

\author{D. Benisty, E. Guendelman, A. Kaganovich, E. Nissimov,and S. Pacheva}

\institute{David Benisty \at Physics Department, Ben-Gurion University of the Negev, 
Beer-Sheva 84105, Israel\\
Frankfurt Institute for Advanced Studies (FIAS), Ruth-Moufang-Strasse~1, 
60438 Frankfurt am Main, Germany\\
\email{benidav@post.bgu.ac.il}
\and Eduardo Guendelman  \at Physics Department, Ben-Gurion University of the Negev, 
Beer-Sheva 84105, Israel\\
Frankfurt Institute for Advanced Studies (FIAS), Ruth-Moufang-Strasse~1, 
60438 Frankfurt am Main, Germany\\ 
Bahamas Advanced Study Institute and Conferences, 4A Ocean Heights, Hill View Circle, 
Stella Maris, Long Island, The Bahamas\\
\email{guendel@bgu.ac.il, guendelman@fias.uni-frankfurt.de}
\and Alexander Kaganovich \at Physics Department, Ben-Gurion University of the Negev,
Beer-Sheva 84105, Israel\\
\email{alexk@bgu.ac.il}
\and Emil Nissimov and Svetlana Pacheva
\at Institute for Nuclear Research and Nuclear Energy,
Bulgarian Academy of Sciences, Sofia, Bulgaria \\
\email{nissimov@inrne.bas.bg, svetlana@inrne.bas.bg}}

\maketitle

\vspace{-0.5in}

\abstract{ 
We present a concise description of the basic features of gravity-matter models 
based on the formalism of non-canonical spacetime volume-forms in its two versions: 
(a) the {\em method of non-Riemannian volume-forms} (metric-independent covariant 
volume elements) and (b) the {\em dynamical spacetime formalism}. Among the principal 
outcomes we briefly discuss: (i) quintessential universe evolution with a 
gravity-``inflaton''-assisted suppression in the ``early'' universe and, 
respectively, dynamical generation in the ``late'' universe of Higgs spontaneous 
electroweak gauge symmetry breaking; (ii) unified description of 
dark energy and dark matter as manifestations of a single material entity -- 
a second scalar field ``darkon''; (iii)unification of dark energy and dark matter
with diffusive interaction among them; (iv) explicit derivation of a stable 
``emergent universe'' solution, \textsl{i.e.}, a creation without Big Bang; 
(v) mechanism for suppression of 5-th force without fine-tuning.}

\section{Introduction -- Non-Riemannian Volume-Form Formalism}

Extended (modified) gravity theories as alternatives/generalizations of the standard
Einstein General Relativity (for detailed accounts, see 
Refs. \ct{extended-grav}-\ct{odintsov-2}) 
are being widely studied in the last decade or so due to pressing motivation 
from cosmology (problems of dark energy and dark matter), quantum field theory 
in curved spacetime (renormalization in higher loops) and string theory
(low-energy effective field theories).

A broad class of actively developed modified/extended gravitational theories is 
based on  employing alternative non-Riemannian spacetime volume-forms 
(metric-independent generally covariant volume elements) in the pertinent 
Lagrangian actions instead of the canonical Riemannian 
one given by the square-root of the determinant of the Riemannian metric 
(originally proposed in \ct{TMT-orig-1,TMT-orig-2}, 
for a concise geometric formulation, see \ct{susyssb-1,grav-bags}).
A characteristic feature of these extended gravitational theories is that
when starting in the first-order (Palatini) formalism the non-Riemannian
volume-forms are almost {\em pure-gauge} degrees of freedom, \textsl{i.e.} 
they {\em do not} introduce any additional propagating gravitational degrees of 
freedom except for few discrete degrees of freedom appearing as
arbitrary integration constants (for a canonical Hamiltonian treatment, see
Appendices A in Refs.\ct{grav-bags,grf-essay}).

Let us recall that volume-forms 
in integrals over differentiable manifolds (not necessarily Riemannian one,
so {\em no} metric is needed) are given by nonsingular maximal rank differential 
forms $\om$:
\br
\int_{\cM} \om \bigl(\ldots\bigr) = \int_{\cM} dx^D\, \Omega \bigl(\ldots\bigr)
\; , \phantom{aaaaaaaa}
\nonu \\
\om = \frac{1}{D!}\om_{\m_1 \ldots \m_D} dx^{\m_1}\wedge \ldots \wedge dx^{\m_D}
\quad ,\quad 
\om_{\m_1 \ldots \m_D} = - \vareps_{\m_1 \ldots \m_D} \Omega \; ,
\lab{omega-1}
\er
(our conventions for the alternating symbols $\vareps^{\m_1,\ldots,\m_D}$ and
$\vareps_{\m_1,\ldots,\m_D}$ are: $\vareps^{01\ldots D-1}=1$ and
$\vareps_{01\ldots D-1}=-1$).
The volume element density (integration measure density) 
$\Omega$ transforms as scalar
density under general coordinate reparametrizations.

In standard generally-covariant theories (with action $S=\int d^D\! x \sqrt{-g} \cL$)
the Riemannian spacetime volume-form is defined through the ``D-bein''
(frame-bundle) canonical one-forms $e^A = e^A_\m dx^\m$ ($A=0,\ldots ,D-1$):
\br
\om = e^0 \wedge \ldots \wedge e^{D-1} = \det\Vert e^A_\m \Vert\,
dx^{\m_1}\wedge \ldots \wedge dx^{\m_D} 
\nonu \\
\longrightarrow \quad
\Omega = \det\Vert e^A_\m \Vert\, d^D x = \sqrt{-\det\Vert g_{\m\n}\Vert}\, d^D x \; .
\lab{omega-riemannian}
\er

Instead of $\sqrt{-g} d^D x$ we can employ another alternative {\em non-Riemannian} 
volume element as in \rf{omega-1} given by a non-singular {\em exact} $D$-form 
$\om = d B$ where:
\br
B = \frac{1}{(D-1)!} B_{\m_1\ldots\m_{D-1}}
dx^{\m_1}\wedge\ldots\wedge dx^{\m_{-1}} 
\nonu \\
\longrightarrow \quad  \Omega \equiv \Phi(B) =
\frac{1}{(D-1)!}\vareps^{\m_1\ldots\m_D}\, \pa_{\m_1} B_{\m_2\ldots\m_D} \; .
\lab{Phi-D}
\er
In other words, the non-Riemannian volume element density is defined in terms of
the dual field-strength scalar density of an auxiliary rank $D-1$ tensor gauge field 
$B_{\m_1\ldots\m_{D-1}}$. 

The plan of exposition is as follows. In Section 2 we describe in some
detail the construction and the main properties of extended gravity models,
based on the formalism of non-Riemannian volume elements, coupled to a
scalar ``inflaton'' field driving the cosmological evolution and a second scalar
``darkon'' field responsible for the unification of dark energy and dark
matter, as well as coupled to the bosonic sector of the standard electorweak
particle model, thus exhibiting a gravity-assisted dynamical generation of
the Higgs electorweak spontaneous symmetry breaking in the post-inflationary
universe. In particular, we find an ``emergent- universe'' cosmological
solution without Big-Bang singularity (on classical level).

Further, in Section 3 we briefly present an alternative mechanism of dark
energy - dark matter unification with diffusive interaction among them based
on the formalism of ``dynamical spacetime'' \ct{Guendelman:2009ck,Benisty:2016ybt}.
Section 4 provides a short discussion of the principal new features which
arise upon inclusion of fermionic fields in modified gravity models based on
the formalism of non-canonical spacetime volume elements as well as on the
requirement of global scale invariance, first of all -- a plausible solution
of the problem of ``fifth force'' without fine-tuning \ct{GK3,GK4}.
The last Section contains our conclusions.

\section{Modified Gravity-Matter Models with Non-Riemannian Volume-Forms --
Cosmological Implications}

To illustrate the main interesting properties of the new class of extended
gravity-matter models based on the non-Riemannian volume-form formalism
we will consider modified gravity in the Palatini formalism coupled 
in a non-standard way via non-Riemannian volume elements to
\ct{grf-essay,BJP-3rd-congress,varna-17}: (i) scalar ``inflaton'' 
field $\vp$; (ii) a second scalar ``darkon'' field $u$; (iii) the bosonic 
fields of the standard electroweak particle model -- 
$\s \equiv (\s_a)$ being a complex $S\!U(2)\times U(1)$ iso-doublet Higgs-like scalar, 
and the $S\!U(2)\times U(1)$ gauge fields $\vec{\cA}_{\m}, \cB_{\m}$.

The ``inflaton'' $\vp$ apart from driving the cosmological evolution triggers
suppression, respectively, generation of the electroweak (Higgs)
spontaneous symmetry breaking in the ``early'', respectively, in the
``late'' universe. The ``darkon'' $u$ is responsible for the unified description of
dark energy and dark matter in the ``late'' universe.

The corresponding action
reads (for simplicity we use units with the Newton constant $G_N = 1/16\pi$):
\br
S = \int d^4 x\,\P_1(A) \Bigl\lb R + L^{(1)}(\vp,\s)\Bigr\rb
\nonu \\
+ \int d^4 x\,\P_2(B) \Bigl\lb L^{(2)}(\vp, \vec{\cA},\cB) 
+ \frac{\P_4 (H)}{\sqrt{-g}}\Bigr\rb 
\nonu \\
- \int d^4 x\,\bigl(\sqrt{-g}+\P_3(C)\bigr) \h g^{\m\n} \pa_\m u \pa_\n u \; .
\lab{TMMT-1}
\er
Here the following notations are used:

(i) $\P_1(A), \P_2(B), \P_3(C)$ are three independent non-Riemannian volume
elements as in \rf{Phi-D} for $D=4$; $\P_4(H)$ is again of the form \rf{Phi-D} 
for $D=4$ and it is needed for consistency of \rf{TMMT-1}.

(ii)The scalar curvature $R$ in Palatini formalism is $R=g^{\m\n} R_{\m\n}(\G)$, 
where the Ricci tensor is a function of the affine connection $\G^\l_{\m\n}$ 
{\em apriori} independent of $g_{\m\n}$.

(iii) The matter field Lagrangians are:
\br
L^{(1)}(\vp,\s) \equiv - \h g^{\m\n} \pa_\m \vp \pa_\n \vp 
- f_1 e^{-\a\vp}
\nonu \\
- g^{\m\n}(\nabla_\m \s)^{*}_a \nabla_\n \s_a
-  \frac{\l}{4} \((\s_a)^{*}\s_a - \m^2\)^2  \; ,
\lab{L-1}
\er
\br
L^{(2)}(\vp, \vec{\cA},\cB) =
-\frac{b}{2} e^{-\a\vp} g^{\m\n} \pa_\m \vp \pa_\n \vp + f_2 e^{-2\a\vp} 
- \frac{1}{4g^2} F^2(\vec{\cA}) - \frac{1}{4g^{\pr\,2}} F^2(\cB) \; ,
\lab{L-2}
\er
where $\a, f_1, f_2$ are dimensionful positive parameters, whereas $b$ is a
dimensionless one ($b$ is needed to obtain a stable ``emergent'' universe
solution, see below \rf{stable-emergent}.
$F^2(\vec{\cA})$ and $F^2(\cB)$ in \rf{L-2} are the 
squares of the field-strengths of the electroweak gauge fields,
and the last term in \rf{L-1} is of the same form as the 
standard Higgs potential.


Let us note that the form of the ``inflaton'' part of the action \rf{TMMT-1} 
is fixed by the requirement of invariance under global Weyl-scale transformations:
\br
g_{\m\n} \to \l g_{\m\n} \;,\; \G^\m_{\n\l} \to \G^\m_{\n\l} \; ,\; 
\vp \to \vp + \frac{1}{\a}\ln \l\; ,
\nonu \\
A_{\m\n\k} \to \l A_{\m\n\k} \; ,\; B_{\m\n\k} \to \l^2 B_{\m\n\k} \; ,\; 
H_{\m\n\k} \to H_{\m\n\k} \; .
\lab{scale-transf}
\er
Scale invariance played an important role in the original papers on the
non-canonical volume-form formalism where also fermions were included 
\ct{TMT-orig-2} (see also Secton 3 below).

The equations of motion of the initial action \rf{TMMT-1} w.r.t. auxiliary
tensor gauge fields $A_{\m\n\l}$, $B_{\m\n\l}$, $C_{\m\n\l}$ and $H_{\m\n\l}$
yield the following algebraic constraints:
\br
R + L^{(1)} =  M_1 = {\rm const} \; ,\; 
L^{(2)} + \frac{\P_4 (H)}{\sqrt{-g}} = - M_2 = {\rm const} \; ,
\nonu
\er
\be
-\h g^{\m\n} \pa_\m u \pa_\n u = M_0 = {\rm const} \; ,\;
\frac{\P_2 (B)}{\sqrt{-g}} \equiv \chi_2 = {\rm const} \; , 
\lab{integr-const}
\ee
where $M_0, M_1, M_2$ are arbitrary dimensionful and $\chi_2$ an
arbitrary dimensionless {\em integration constants}. 

The equations of motion of \rf{TMMT-1} w.r.t. affine connection $\G^\m_{\n\l}$
yield a solution for $\G^\m_{\n\l}$ as a Levi-Civita connection 
$\G^\m_{\n\l} = \G^\m_{\n\l}({\bar g}) =
\h {\bar g}^{\m\k}\(\pa_\n {\bar g}_{\l\k} + \pa_\l {\bar g}_{\n\k}
- \pa_\k {\bar g}_{\n\l}\)$ w.r.t. to the a {\em Weyl-rescaled metric} 
${\bar g}_{\m\n} = \chi_1 g_{\m\n} \; ,\; \chi_1 \equiv \frac{\P_1 (A)}{\sqrt{-g}}$.

The passage to the ``Einstein-frame'' (EF) is accomplished by a Weyl-conformal
transformation to ${\bar g}_{\m\n}$ upon using relations
\rf{integr-const}, so that the EF action with a canonical Hilbert-Einstein gravity
part w.r.t. ${\bar g}_{\m\n}$ and with the canonical Riemannian
volume element density $\sqrt{\det\vert\vert -{\bar g}_{\m\n}\vert\vert}$ reads:
\be
S_{\rm EF} = \int d^4 x \sqrt{-{\bar g}} \Bigl\lb R({\bar g}) + L_{\rm EF}\Bigr\rb \; ,
\lab{EF-action}
\ee
and where the EF matter Lagrangian turns out to be of a
quadratic ``k-essence'' type \ct{k-essence-1}-\ct{k-essence-4}
w.r.t. both the ``inflaton'' $\vp$ and ``darkon'' $u$ fields:
\br
L_{\rm EF} = {\bar X}
- {\bar Y}\Bigl\lb f_1 e^{-\a\vp} + \frac{\l}{4} \((\s_a)^{*}\s_a - \m^2\)^2 + M_1 
- \chi_2 b e^{-\a\vp}{\bar X}\Bigr\rb 
\nonu \\
+ {\bar Y}^2 \Bigl\lb \chi_2 (f_2 e^{-2\a\vp} + M_2) + M_0\Bigr\rb 
+ L\lb\s,\vec{\cA},\cB\rb \; ,
\lab{L-eff-total}
\er
with $L\lb\s,\vec{\cA},\cB\rb \equiv
- {\bar g}^{\m\n} (\nabla_\m \s_a)^{*}\nabla_\n \s_a
- \frac{\chi_2}{4g^2} {\bar F}^2(\vec{\cA}) - 
\frac{\chi_2}{4g^{\pr\,2}} {\bar F}^2(\cB)$. In \rf{L-eff-total} all quantities 
defined in terms of the EF metric ${\bar g}_{\m\n}$ are indicated by an upper bar, 
and the following short-hand notations are used: 
${\bar X} \equiv - \h {\bar g}^{\m\n} \pa_\m \vp \pa_\n \vp \;,\;
{\bar Y} \equiv - \h{\bar g}^{\m\n}\pa_\m u \pa_\n u$.

From \rf{L-eff-total} we deduce the following full effective scalar potential:
\be
U_{\rm eff}\bigl(\vp,\s\bigr) = 
\frac{\Bigl(f_1 e^{-\a\vp} + \frac{\l}{4} \((\s_a)^{*}\s_a - \m^2\)^2 
+ M_1\Bigr)^2}{4\bigl\lb \chi_2 (f_2 e^{-2\a\vp} + M_2) + M_0\bigr\rb}
\lab{U-eff-total}
\ee

As discussed in Refs.\ct{BJP-3rd-congress,varna-17} 
$U_{\rm eff}(\vp,\s)$ \rf{U-eff-total} has few remarkable properties.
First, $U_{\rm eff}(\vp,\s)$ possesses two infinitely large flat regions
as function of $\vp$ when $\s$ is fixed:

(a) (-) flat ``inflaton'' region for large negative values of $\vp$
corresponding to the evolution of the ``early'' universe;

(b) (+) flat ``inflaton'' region for large positive values of $\vp$ with
$\s$ fixed corresponding to the evolution of the ``late'' universe''.

This is graphically depicted on Fig.1.

\begin{figure}
\begin{center}
\includegraphics[width=9cm,keepaspectratio=true]{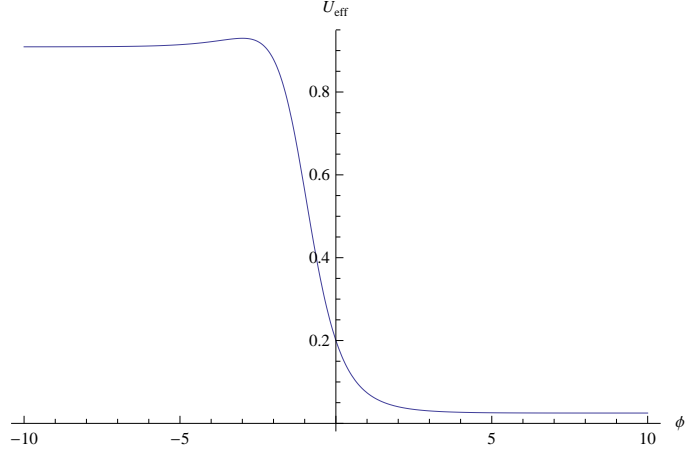}
\caption{Qualitative shape of the effective scalar potential $U_{\rm eff}$
\rf{U-eff-total} as function of the ``inflaton'' $\vp$ for $M_1 > 0$ and
fixed Higgs-like $\s$}
\end{center}
\end{figure}

In the (-) flat ``inflaton'' region, \textsl{i.e.}, in the ``early'' universe
the effective scalar field potential \rf{U-eff-total} reduces to (an
aproximately) constant value
\be
U_{\rm eff}\bigl(\vp,\s\bigr) \simeq U_{(-)} 
= \frac{f_1^2}{4\chi_2\,f_2}
\lab{U-minus}
\ee
Thus, there is no $\s$-field potential and, therefore,
{\em no electroweak spontaneous breakdown in the ``early'' universe}.

On the other hand, in the (+) flat ``inflaton'' region, \textsl{i.e.}, in the
``late'' universe the effective scalar field potential becomes:
\be
U_{\rm eff}\bigl(\vp,\s\bigr) \simeq U_{(+)}(\s) =
\frac{\Bigl(\frac{\l}{4} \((\s_a)^{*}\s_a - \m^2\)^2 + M_1\Bigr)^2}{
4\bigl(\chi_2 M_2 + M_0\bigr)} \; ,
\lab{U-plus-higgs}
\ee
which obviously yields {\em nontrivial vacuum for the Higgs-like field} 
$|\s_{\rm vac}|= \m$. Therefore, in the ``late'' universe we have the
standard spontaneous breakdown of electroweak $SU(2)\times U(1)$ ~gauge symmetry.
Moreover, at the Higgs vacuum we obtain from \rf{U-plus-higgs}
a dynamically generated cosmological constant $\L_{(+)}$ of the ``late'' Universe:
\be
U_{(+)}(\m) \equiv 2\L_{(+)} = \frac{M_1^2}{4\bigl(\chi_2 M_2 + M_0\bigr)} \; .
\lab{CC-eff-plus}
\ee 
If we identify  the integration constants with the
fundamental scales in Nature as $M_{0,1} \sim M^4_{EW}$ and $M_2 \sim M^4_{Pl}$, where
where $M_{\rm Pl}$ is the Planck mass scale and 
$M_{\rm EW} \sim 10^{-16} M_{\rm Pl}$ is the electroweak mass scale, then
$\L_{(+)}\sim M^8_{EW}/M^4_{Pl} \sim 10^{-120} M^4_{Pl}$ ,
which is the right order of magnitude for the present epoch's vacuum energy
density as already realized in \ct{arkani-hamed}.

On the other hand, if we take the order of magnitude of the coupling
constants in the effective potential \rf{U-eff-total}
$f_1 \sim f_2 \sim (10^{-2} M_{Pl})^4$, 
then the order of magnitude of the vacuum energy density of the ``early'' universe 
\rf{U-minus} becomes:
\be
U_{(-)} \sim f_1^2/f_2 \sim 10^{-8} M_{Pl}^4 \; ,
\lab{U-minus-magnitude}
\ee
which conforms to the Planck Collaboration data \ct{Planck-1,Planck-2} 
implying the energy scale of inflation of order $10^{-2} M_{Pl}$.

Now, let us perform FLRW reduction of the EF action \rf{EF-action}.
\textsl{i.e.}, restricting the metric ${\bar g}_{\m\n}$ to the FLRW form
$ds^2 = {\bar g}_{\m\n} dx^\m dx^\n = - dt^2 + a^2(t) d{\vec x}^2$. Thus
we obtain in
the ``late'' universe, \textsl{i.e.}, for large positive ``inflaton'' $\vp$
values the following results for the density, pressure, the Friedmann scale
factor (the solution for $a(t)$ below first appeared in \ct{turner-etal})
and the ``inflaton'' velocity:
\br
\rho = \frac{M_1^2}{4(\chi_2 M_2 + M_0)} + \frac{\pi_u}{a^3}\,
\Bigl\lb\frac{M_1}{\chi_2 M_2 + M_0}\Bigr\rb^{\h} 
+ {\rm O} \bigl(\frac{\pi^2_u}{a^6}\bigr) \; ,
\lab{rho-plus} \\
p = - \frac{M_1^2}{4(\chi_2 M_2 + M_0)} + 
{\rm O} \bigl(\frac{\pi^2_u}{a^6}\bigr) \; ,
\lab{p-plus}
\er
\br
a(t) \simeq \Bigl(\frac{C_0}{2\L_{(+)}}\Bigr)^{1/3} 
\sinh^{2/3}\Bigl(\sqrt{\frac{3}{4}\L_{(+)}}\, t\Bigr) \; ,
\lab{a-plus} \\
\vpdot \simeq {\rm const}\, \sinh^{-2}\Bigl(\sqrt{\frac{3}{4}\L_{(+)}}\, t\Bigr) \; ,
\lab{vpdot-plus}
\er
where $\pi_u$ is the conserved ``darkon'' canonical momentum, $\L_{(+)}$ is as in 
\rf{CC-eff-plus} and $C_0 \equiv \pi_u \sqrt{M_1 (\chi_2 M_2 + M_0)^{-1}}$.

Relations \rf{rho-plus}-\rf{p-plus} straightforwardly show that in the ``late''
universe we have explicit unification of dark energy (given by
the dynamically generated cosmological constant \rf{CC-eff-plus} -- first
constant terms on the r.h.sides in \rf{rho-plus} and \rf{p-plus}, 
and dark matter given as a ``dust'' fluid contribution -- second term 
${\rm O}(a^{-3})$ on the r.h.s. of \rf{rho-plus}.

A further interesting property under consideration is the existence of a stable 
``emergent'' universe solution -- a creation without Big Bang (cf. 
Refs.\ct{emergent,cuba}). It is characterized by the condition on the Hubble 
parameter $H$:
\br
H=0 \quad \to \quad a(t) = a_0 = {\rm const} \;\; , \;\;\rho + 3p = 0 \; ,
\nonu \\
\frac{K}{a_0^2} = \frac{1}{6}\rho ~(= {\rm const}) \; , \phantom{aaaaaaaa}
\lab{emergent-cond}
\er
and the ``inflaton'' is on the $(-)$ flat region (large negative values of $\vp$). Then
relations \rf{emergent-cond} together with the ``inflaton'' and ``darkon''
equations of motion imply that also ``inflaton'' velocity $\vpdot = {\rm const}$
and the constant density and pressure read:
\br
\rho \simeq - \frac{3\chi_2 b^2}{16 f_2} \vpdot^4
\h \vpdot^2 \Bigl(1+\frac{bf_1}{2f_2}\Bigr) + \frac{f_1^2}{4\chi_2 f_2} \; ,
\lab{rho-0} \\
p \simeq - \frac{\chi_2 b^2}{16 f_2} \vpdot^4
\h \vpdot^2 \Bigl(1+\frac{bf_1}{2 f_2}\Bigr) - \frac{f_1^2}{4\chi_2 f_2} \; .
\lab{p-0}
\er
The truncated Friedmann Eqs.\rf{emergent-cond} yield exact solutions for the
constant ``inflaton'' velocity $\vpdot_0$ and Friedmann factor $a_0$:
\be
\vpdot_0^2 = \frac{8 f_2}{3\chi_2 b^2} \Bigl\lb 1+\frac{bf_1}{2f_2} 
- \sqrt{\bigl(1+\frac{bf_1}{2f_2}\bigr)^2 - \frac{3b^2 f_1^2}{16 f_2^2}}\Bigr\rb \; ,
\lab{vpdot-a-const} 
\ee
and $a_0^2 = 6K/\rho_0$ where:
\br
\rho_0 = \frac{f_1^2}{2\chi_2 f_2} 
- \h \vpdot_0^2 \Bigl(1+\frac{bf_1}{2f_2}\Bigr) \; .
\lab{rho-00}
\er
Studying perturbation $a \to a + \d a(t)$ of the ``emergent'' universe
condition \rf{emergent-cond} we obtain a harmonic oscillator equation for 
$\d a(t)$ (here $\vpdot_0^2$ as in \rf{vpdot-a-const}, and $\rho_0$ as in
\rf{rho-00}):
\br
\d\addot + \om^2 \d a =0 \;, \phantom{aaaaaaaa}
\nonu\\
\om^2 \equiv\frac{\rho_0}{6}\Bigl\lb 3\frac{\h(1+bf_1/2f_2) - 
\vpdot_0^2 \chi_2 b^2/8f_2}{\vpdot_0^2 3\chi_2 b^2/8f_2 - \h(1+bf_1/2f_2)}
-1\Bigr\rb > 0 
\lab{stable-emergent}
\er
for ~$ - 8(1-\h\sqrt{3})\frac{f_2}{f_1} < b < - \frac{f_2}{f_1}$ .

The non-Riemannian volume-form formalism was also successfully applied to
propose an qualitatively new mechanism for a {\em dynamical spontaneous breaking of
supersymmetry} in supergravity by constructing modified formulation of 
standard minimal $N=1$ supergravity as well as of anti-de Sitter supergravity in terms 
of a non-Riemannian volume elements \ct{susyssb-1,susyssb-2}. 
This naturally triggers the appearance of a 
{\em dynamically generated cosmological constant} as an arbitrary integration 
constant which signifies {\em dynamical spontaneous supersymmetry breakdown}.
The same formalism applied to anti-de Sitter supergravity allows us to appropriately 
choose the above mentioned arbitrary integration constant so as to obtain 
simultaneously a {\em very small effective observable cosmological constant} as well
as a {\em large physical gravitino mass} as required by modern cosmological 
scenarios for slowly expanding universe of the present epoch \ct{slow-accel-1,slow-accel-2,slow-accel-3}.

\section{Dynamical spacetime formulation} 

Let us now observe that the non-Riemannian volume element density $\Omega=\Phi(B)$
\rf{Phi-D} on a Riemannian manifold can be rewritten using Hodge duality
(here $D=4$) in terms of a vector field
$\chi^\m = \frac{1}{3!} \frac{1}{\sqrt{-g}} \vareps^{\m\n\k\l} B_{\n\k\l}$
so that $\Omega$ becomes $\Omega (\chi) = \pa_\m \bigl(\sqrt{-g}\chi^\m\bigr)$,
\textsl{i.e.} it is a non-canonical volume element density different from $\sqrt{-g}$, 
but involving the metric. It can be represented alternatively through a
Lagrangian multiplier action term yielding covariant conservation of a
specific energy-momentum tensor of the form $\cT^{\m\n} = g^{\m\n} \cL$:
\be
\mathcal{S}_{(\chi)}=\int d^{4}x\sqrt{-g} \, \chi_{\mu;\nu}\cT^{\mu\nu}
= \int d^4 x \pa_\m \bigl(\sqrt{-g}\chi^\m\bigr) \bigl(-\cL\bigr) \; ,
\lab{action}
\ee
where $\chi_{\m;\n}=\pa_\n\chi_\m -\G_{\m\n}^\l\chi_\l$. 

 
The vector field $\chi_\mu$ is called {\em ``dynamical space time vector''}
, because of the energy density of $\cT^{00}$ is a canonically conjugated momentum 
w.r.t. $\chi_0$, which is what we expected from a dynamical time. 

In what follows we will briefly consider a new class of gravity-matter
theories based on the ordinary Riemannian volume element density $\sqrt{-g}$ but
involving action terms of the form \rf{action} where now $\cT^{\m\n}$ is of
more general form than $\cT^{\m\n} = g^{\m\n} \cL$. 
This new formalism is called {\em ``dynamical spacetime formalism''} 
\ct{Guendelman:2009ck,Benisty:2016ybt} due to the above remark on $\chi_0$. 

Ref.\cite{Benisty:2018qed} describes a unification between dark energy and dark 
matter by introducing a quintessential scalar field in addition to the dynamical 
time action. The total Lagrangian reads:
\be
\mathcal{L}=\frac{1}{2}R+\chi_{\mu;\nu}\cT^{\mu\nu} - 
\frac{1}{2}g^{\alpha\beta} \phi_{,\alpha}\phi_{,\beta} - V(\phi),
\lab{action-with-xi}
\ee
with energy-momentum tensor 
$\cT^{\mu\nu} = -\frac{1}{2} \phi^{,\mu} \phi^{,\nu}$. 
From the variation of the Lagrangian term $\chi_{\mu;\nu}\cT^{\mu\nu}$ with respect 
to the vector field $\chi_{\mu}$, the covariant conservation of the energy-momentum 
tensor $\nabla_{\mu} \cT^{\m\n} = 0$ is implemented. The latter within the FLRW
framework forces the kinetic term of the scalar field to behave
as a dark matter component:
\begin{equation}\label{var1}
\nabla_{\mu}\cT^{\mu\nu} = 0 \quad \Rightarrow \quad \dot{\phi}^2 = \frac{2 \Omega_{m0}}{a^3}.
\end{equation}
where $\Omega_{m0}$ is an integration constant. 
The variation with respect to the scalar field $\phi$ yields a current: 
\begin{equation}\label{var22}
- V'(\phi) = \nabla_\mu j^{\mu} , \quad j^{\mu}  = \frac{1}{2}\phi_{,\nu} (\chi^{\mu;\nu}+\chi^{\nu;\mu}) + \phi^{,\mu}
\end{equation}
For constant potential $V(\phi) = \Omega_\Lambda = const$ the current is 
covariantly conserved. 

In the FLRW setting, where the dynamical time
ansatz introduces only a time component $\chi_\mu =(\chi_0,0,0,0)$, 
the variation (\ref{var22}) gives:
\begin{equation}\label{var2}
 \dot{\chi}_0 - 1 = \xi \,  a^{-3/2}  ,
\end{equation}
where $\xi$ is an integration constant. Accordingly, the FLRW energy density and
pressure read:
\begin{equation}
\label{setvv}
\rho = \left(\dot{\chi}_0-\frac{1}{2}\right)\dot{\phi}^2 + V, \quad p = \frac{1}{2}\dot{\phi}^2(\dot{\chi}_0-1) - V.
\end{equation}
Plugging the relations (\ref{var1},\ref{var2}) into the density and the pressure 
terms (\ref{setvv}) yields the following simple form of the latter:
\be
\rho = \Omega_\Lambda +\frac{\xi \Omega_{m 0}}{a^{9/2}} +\frac{\Omega_{m 0}}{a^3}, \quad p = -\Omega_\Lambda+ \frac{\xi \Omega_{m0}}{2\, a^{9/2}}.
\lab{Frid}
\ee
In \rf{Frid} there are 3 components for the "dark fluid": dark energy with 
$\omega_\Lambda = -1$, dark matter with $\omega_m = 0$ and an additional equation 
of state $\omega_\xi = 1/2$.
For non-vanishing and negative $\xi$ the additional part introduces a minimal 
scale parameter, which avoids singularities. If the dynamical time is equivalent 
to the cosmic time $\chi_0 = t$, we obtain $\xi = 0$ from Eq.(\ref{var2}), 
whereupon the density and the pressure terms \rf{Frid} coincide with those from
the $\Lambda$CDM model precisely. 
The additional part (for $\xi \neq 0$) fits more to the late time accelerated 
expansion data, as observed in Ref. \cite{Anagnostopoulos:2019myt}.

Ref. \cite{Benisty:2018gzx} shows that with higher dimensions, the solution 
derived from the Lagrangian \rf{action-with-xi} describes inflation, where the total
volume oscillates and the original scale parameter exponentially {\em grows}. 

The dynamical spacetime Lagrangian can be generalized to yield  a 
{\em diffusive energy-momentum tensor}. Ref. \cite{Calogero:2013zba} shows that
the diffusion equation has the form:
\begin{equation}\label{diffusion}
\nabla_\mu \cT^{\mu\nu}=3\sigma j^\nu, \quad j^{\mu}_{;\mu}=0, \end{equation}
where $\sigma$ is the diffusion coefficient and $j^{\mu}$ is a current source. 
The covariant conservation of the current source indicates the conservation of 
the number of the particles. 
By introducing the vector field $\chi_\mu$ in a different part of the 
Lagrangian:
\begin{equation} \label{nhd1}
\mathcal{L}_{(\chi,A)}=\chi_{\mu;\nu}\cT^{\mu\nu} +\frac{\sigma}{2} (\chi_{\mu}+\partial_{\mu}A)^2, 
\end{equation} 
the energy-momentum tensor $\cT^{\mu\nu}$ {\em gets a diffusive source}.
From a variation with respect to the dynamical 
space time vector field $\chi_{\mu}$ we obtain:
\begin{equation} \label{nhd2}
\nabla_{\nu}\cT^{\mu\nu}=\sigma(\chi^{\mu}+\partial^{\mu}A)= f^\mu,
\end{equation} 
a current source $f^\mu=\sigma (\chi^{\mu}+\partial^{\mu} A )$ for the 
energy-momentum tensor. From the variation with respect to the new scalar $A$,
a covariant conservation of the current emerges $f^\mu_{;\mu}=0$. 
{\em The latter relations correspond to the diffusion equation (\ref{diffusion})}.

Refs.\cite{Benisty:2017eqh,Benisty:2017rbw,Benisty:2017lmt,Bahamonde:2018uao} 
study the cosmological solution using the energy-momentum tensor
$\cT^{\mu\nu} = -\frac{1}{2}g^{\mu\nu} \phi^{,\lambda}\phi_{\lambda}$. 
The total Lagrangian reads:
\begin{equation}
\mathcal{L}=\frac{1}{2}R - \frac{1}{2}g^{\alpha\beta} \phi_{,\alpha}\phi_{,\beta} - V(\phi) +\chi_{\mu;\nu}\cT^{\mu\nu} +\frac{\sigma}{2} (\chi_{\mu}+\partial_{\mu}A)^2.
\end{equation}
The FLRW solution unifies the dark energy and the dark matter originating from 
one scalar field with possible diffusion interaction. 
Ref.\cite{Benisty:2018oyy} investigates more general energy-momentum tensor 
combinations and shows that asymptotically all of the combinations yield 
$\Lambda$CDM model as a stable fixed point.

\section{Scale Invariance, Fifth Force and Fermionic Matter} 

The originally proposed theory with two volume element densities
(integration measure densities) \ct{TMT-orig-1,TMT-orig-2}, where at least one of 
them was a non-canonical one and short-termed ``two-measure theory'' (TMT), 
has a number of remarkable properties 
if fermions are included in a self-consistent way \ct{TMT-orig-2}.
In this case, the constraint that arises in the TMT models in the Palatini 
formalism can be represented as an equation for $\chi\equiv\Phi/\sqrt{-g}$, 
in which the left side has an order of the vacuum energy density, 
and the right side (in the case of non-relativistic fermions) is proportional 
to the fermion density. 
Moreover, it turns out that even cold fermions have a (non-canonical) 
pressure $P_f^{noncan}$ and the corresponding contribution to the energy-momentum 
tensor has the structure of a  cosmological constant term which is proportional 
to the fermion density. The  remarkable fact is that the right hand side of the 
constraint coincide with $P_f^{noncan}$. This allows us to construct 
 a cosmological model\cite{GK1} of the late universe in which 
dark energy is generated by a gas of non-relativistic neutrinos without the need 
to introduce into the model a specially designed scalar field. 

In models with a scalar field, the requirement of scale invariance of the 
initial action\cite{TMT-orig-1} plays a very constructive role. 
It allows to construct a model\cite{GK2} where without 
fine tuning we have realized: absence of initial singularity of the curvature; 
k-essence;  inflation with graceful exit to zero cosmological constant.

Of particular interest are scale invariant  models in which both fermions and 
a dilaton scalar field $\phi$ are present. Then it turns out that the Yukawa 
coupling of fermions to $\phi$ is proportional to  $P_f^{noncan}$. As a result, 
it follows from the constraint, that in all cases when fermions are in states 
which constitute a regular barionic matter, the Yukawa coupling of fermions to 
dilaton has an order of ratio of the vacuum energy density to the fermion energy 
density\cite{GK3}. Thus, the theory provides 
a solution of the 5-th force problem without any fine tuning or a special design 
of the model. Besides, in the described states, the regular Enstein's equations 
are reproduced. 
In the opposite case, when fermions are very deluted, e.g. in the model of the 
late Universe filled with a cold neutrino gas, the neutrino dark energy appears 
in such a way that the dilaton $\phi$ dynamics is closely correlated with that of the 
neutrino gas\cite{GK3}.  

A scale invariant model
containing a dilaton $\phi$ and dust (as a
model of matter)\cite{GK4} possesses similar features. The
dilaton to matter coupling "constant" $f$ appears to be dependent
of the matter density. In normal conditions, i.e. when the matter
energy density is many orders of magnitude larger than the dilaton
contribution to the dark energy density, $f$ becomes less than the
ratio of the "mass of the vacuum" in the volume occupied by the
matter to the Planck mass. The model yields this kind of
"Archimedes law" without any especial (intended for this) choice
of the underlying action and without fine tuning of the
parameters. The model not only explains why all attempts to
discover a scalar force correction to Newtonian gravity were
unsuccessful so far but also predicts that in the near future
there is no chance to detect such corrections  in the astronomical
measurements as well as in the specially designed  fifth force
experiments on intermediate, short (like millimeter) and even
ultrashort (a few nanometer) ranges. This prediction is
alternative to predictions of other known models.

More recently other authors have rediscovered the important role of 
scale invariance in the avoidance of a 5-th force \cite{Hill-etal}.
We should point out that our original work \ct{GK3,GK4} on avoidance of the 
5-th force through scale invariance symmetry preceeds that of 
Ref.\cite{Hill-etal} by a substantial number of years.


\section{Conclusions}
In the present paper we describe in some details the principal physically
interesting features of a specific class on extended (modified)
gravitational theories beyong the standard Einstein's general relativity.
They are constructed in terms of non-Riemannian spacetime volume forms
(metric-independent non-canonical volume elements). An important role is
also being played by the requirement of global scale invariance. We present
a modified gravity-matter model where gravity is coupled in a non-canonical
way to two scalar fields (``inflaton'' and ``darkon'') as well as to the
bosonic sector of the standard electroweak model of elementary particle
physics. The ``inflaton'' scalar field triggers a quintessential inflationary 
evolution of the Universe where all energy scales are determined dynamically 
through free integration constants arising due to the modified gravitational dynamics 
because of the non-Riemannian volume elements. The ``darkon'' scalar field on its
part creates through its dynamics a unified description of dark energy and
dark matter. A particularly notable feature is the
gravity-''inflaton''-assisted dynamical generation of Higgs electroweak 
spontaneous symmetry breaking in the post-inflationary epoch and its
suppression in the ealy-universe stage. 
Under special initial condition on the Hubble parameter we find
(on classical level) an ``emergent universe'' solution describing early
universe evolution without spacetime singularities (no ``Big Bang'').

Furthermore, we have briefly discussed a parallel alternative non-canonical
spacetime volume element approach based on the concept of ``dynamical
spacetime'' and have demonstrated the appearance of unified description of
dark energy and dark matter with a diffusive interaction among them. Finally
we briefly outlined, based on our original work \ct{GK3,GK4},
how the formalism of non-canonical volume elements in
modified gravity-matter models with fermions provides a resolution of the
problem of ``fifth force'' without any fine tunings.

In the above constructions we have employed the first-order (Palatini)
formalism in the initial gravity actions. Further physically interesting
features are obtained when combining the non-Riemannian spacetime volume
element formalism with the second order (metric) gravity formalism. In
particular, in the latter case it was recently shown \ct{dyn-infl} 
that starting with a
pure modified gravity in terms of several non-Riemannian volume elements and without
any initial matter fields one creates dynamically (in the ``Einstein
frame'') a canonical scalar field with a non-trivial inflationary potential
generalizing the classical Starobinsky potential \ct{starobinsky} and
yielding results for the cosmological observables (scalar power spectral
index and the tensor-to-scalar ratio) fitting very well to the avaible
observational data \ct{PLANCK}.

\begin{acknowledgement}
We gratefully acknowledge support of our collaboration through 
the academic exchange agreement between the Ben-Gurion University in Beer-Sheva,
Israel, and the Bulgarian Academy of Sciences. 
E.N. and E.G. have received partial support from European COST actions
MP-1405, CA-16104, CA-18108 and from CA-15117, CA-16104, CA-18108 respectively.
E.N. and S.P. are also thankful to Bulgarian National Science Fund for
support via research grant DN-18/1.
\end{acknowledgement}


\begin{thebibliography}{99}
\bibitem{extended-grav}
S. Capozziello and M. De Laurentis, \PRep{509}{2011}{167} ~(\textsl{arXiv:1108.6266}).
\bibitem{extended-grav-book}
S. Capozziello and V. Faraoni, {\em ``Beyond Einstein Gravity -- 
A Survey of Gravitational Theories for Cosmology and Astrophysics''},
(Springer, 2011).
\bibitem{odintsov-1}
S. Nojiri and S. Odintsov, \PRep{505}{2011}{59}.
\bibitem{odintsov-2}
S. Nojiri, S. Odintsov and V. Oikonomou, \PRep{692}{2017}{1}
~(\textsl{arXiv:1705.11098}).
\bibitem{TMT-orig-1}
E. Guendelman, {Mod. Phys. Lett.} {\bf A14}, 1043-1052 (1999)
~(\textsl{arXiv:gr-qc/9901017}).
\bibitem{TMT-orig-2}
E. Guendelman and A. Kaganovich,
{Phys. Rev.} {\bf D60} 065004 (1999) ~(\textsl{arXiv:gr-qc/9905029}).
\bibitem{susyssb-1}
E. Guendelman, E. Nissimov and S. Pacheva, {Bulg. J. Phys.} {\bf 41}, 
123 (2014) ~(\textsl{arXiv:1404.4733}).
\bibitem{grav-bags}
E. Guendelman, E. Nissimov and S. Pacheva, \IJMPA{30}{2015}{1550133} 
~(\textsl{arXiv:1504.01031}).
\bibitem{grf-essay}
E. Guendelman, E. Nissimov and S. Pacheva, \IJMPD{25}{2016}{1644008}
~(\textsl{1603.06231}).
\bibitem{Guendelman:2009ck} 
E.~Guendelman,
\IJMPA{25}{2010}{4081} 
~(\textsl{arXiv:0911.0178 [gr-qc]}).
\bibitem{Benisty:2016ybt} 
D.~Benisty and E.~I.~Guendelman,
\MPLA{31}{2016}{1650188}  
~(\textsl{arXiv:1609.03189 [gr-qc]}).
\bibitem{GK3} 
E. Guendelman and A. Kaganovich, \IJMPA{21}{2006}{4373} 
~(\textsl{arXiv:gr-qc/0603070}).
\bibitem{GK4} 
E. Guendelman and A. Kaganovich, \AoP{323}{2008}{866}
~(\textsl{arXiv:0704.1998}).
\bibitem{BJP-3rd-congress}
E. Guendelman, E. Nissimov and S. Pacheva,  {Bulg. J. Phys.} {\bf 44}, 15-30
(2017) ~(\textsl{arXiv:1609.06915}).
\bibitem{varna-17}
E. Guendelman, E. Nissimov and S. Pacheva, in {\em  Springer Proceedings in 
Mathematics and Statistics} v.255: \textsl{Quantum Theory and Symmetries with 
Lie Theory and Its Applications in Physics, vol.2}, ed. V. Dobrev, pp.99-114 
(Springer, 2018) ~(\textsl{arXiv:1712.09844}).
\bibitem{k-essence-1}
T. Chiba, T.Okabe and M. Yamaguchi, \PRD{62}{2000}{023511}
~(\textsl{arXiv:astro-ph/9912463}).
\bibitem{k-essence-2}
C. Armendariz-Picon, V. Mukhanov and P. Steinhardt, \PRL{85}{2000}{4438}
~(\textsl{arXiv:astro-ph/0004134}).
\bibitem{k-essence-3}
C. Armendariz-Picon, V. Mukhanov and P. Steinhardt,
\PRD{63}{2001}{103510} ~(\textsl{arXiv:astro-ph/0006373}).
\bibitem{k-essence-4}
T. Chiba, \PRD{66}{2002}{063514} ~(\textsl{arXiv:astro-ph/0206298}).
\bibitem{arkani-hamed}
N. Arkani-Hamed, L.J. Hall, C. Kolda and H. Murayama, \PRL{85}{2000}{4434-4437}
~(\textsl{astro-ph/0005111}).
\bibitem{Planck-1}
R. Adam {\it et al.} (Planck Collaboration), {Astron. Astrophys.}
{\bf 571}, A22 (2014) ~(\textsl{arXiv:1303.5082} [astro-ph.CO]).
\bibitem{Planck-2}
R. Adam {\em et al.} (Planck Collaboration), {Astron. Astrophys.}
{\bf 586}, A133 (2016) ~(\textsl{arXiv:1409.5738} [astro-ph.CO]).
\bibitem{turner-etal}
J. Frieman, M. Turner and D. Huterer, \textsl{Ann. Rev. Astron. Astrophys.}
{\bf 46}, 385-432 (2008) ~(\textsl{arXiv:0803.0982}).
\bibitem{susyssb-2}
E. Guendelman, E. Nissimov and S. Pacheva, in: {\em VIII
Mathematical Physics Meeting}, pp. 105-115, ed. by B. Dragovich and I. Salom
(Belgrade Inst. Phys. Press, 2015) ~(\textsl{arXiv:1501.05518}).
\bibitem{emergent}
E. Guendelman, R. Herrera, P. Labrana, E. Nissimov and S. Pacheva,
\textsl{Gen. Rel. Grav.} {\bf 47}, art.10 (2015) ~(\textsl{arXiv:1408.5344v4}).
\bibitem{cuba}
E. Guendelman, R. Herrera, P. Labrana, E. Nissimov and S. Pacheva,
\textsl{Astronomische Nachrichten} {\bf 336}, 810 (2015) 
~(\textsl{arXiv:1507.08878}).
\bibitem{slow-accel-1}
Adam G. Riess {\em et.al.} (Supernova Search Team),
\textsl{Astronomical Journal} {\bf 116} (1998) 1009-1038 
~(\textsl{arxiv:astro-ph/9805201}).
\bibitem{slow-accel-2}
S. Perlmutter  {\em et.al.} (The Supernova Cosmology Project),
\textsl{Astrophysical Journal} {\bf 517} (1999) 565-586 
~(\textsl{arxiv:astro-ph/9812133}).
\bibitem{slow-accel-3}
A.G. Riess {\em et.al.}, \textsl{Astrophysical Journal} {\bf 607} (2004) 665-687
~(\textsl{arxiv:astro-ph/0402512}).
\bibitem{Benisty:2018qed} 
D.~Benisty and E.~Guendelman,
\PRD{98}{2018}{023506} 
~(\textsl{arXiv:1802.07981 [gr-qc]}).
\bibitem{Anagnostopoulos:2019myt} 
F.~Anagnostopoulos, D.~Benisty, S.~Basilakos and E.~Guendelman,
(\textsl{arXiv:1904.05762 [gr-qc]}).
\bibitem{Benisty:2018gzx} 
D.~Benisty and E.~Guendelman,
\PRD{98}{2018}{043522}  
~(\textsl{arXiv:1805.09314 [gr-qc]}).
\bibitem{Calogero:2013zba} 
  S.~Calogero and H.~Velten,
  JCAP {\bf 1311}, 025 (2013)
  doi:10.1088/1475-7516/2013/11/025
  [arXiv:1308.3393 [astro-ph.CO]].
\bibitem{Benisty:2018oyy} 
D.~Benisty, E.~Guendelman and Z.~Haba,
\PRD{99}{2019}{123521}
~\textsl{arXiv:1812.06151 [gr-qc]}.
\bibitem{Benisty:2017eqh} 
D.~Benisty and E.~Guendelman,
Eur. Phys. J. {\bf C77}, 396 (2017)
~(\textsl{arXiv:1701.08667 [gr-qc]}).
\bibitem{Benisty:2017rbw} 
D.~Benisty and E.~Guendelman,
\IJMPD{26}{2017}{1743021} 
\bibitem{Benisty:2017lmt} 
D.~Benisty and E.~Guendelman,
\IJMPA{33}{2018}{1850119}  
~(\textsl{arXiv:1710.10588 [gr-qc]}).
\bibitem{Bahamonde:2018uao} 
S.~Bahamonde, D.~Benisty and E.~Guendelman,
\textsl{Universe} {\bf 4}, 112 (2018)
~(\textsl{arXiv:1801.08334 [gr-qc]}).
\bibitem{GK1}
E. Guendelman and A. Kaganovich, \PRD{87}{2013}{044021}
~(\textsl{arXiv:1208.2132}). 
\bibitem{GK2} 
E. Guendelman and A. Kaganovich, \PRD{75}{2007}{083505} 
~(\textsl{arXiv:gr-qc/0607111}).
\bibitem{dyn-infl} 
D. Benisty, E. Guendelman, E. Nissimov and S. Pacheva, 
\textsl{Eur. Phys. J.} {\bf C79}, 806 (2019) ~(\textsl{arXiv:1906.06691}). 
\bibitem{starobinsky}
A. Starobinsky,
\textsl{JETP Lett.}  {\bf 30} (1979) 682 
[Pisma Zh.\ Eksp.\ Teor.\ Fiz.\  {\bf 30} (1979) 719].
\bibitem{PLANCK}
Y. Akrami {\it et al.} [Planck Collaboration],
~\textsl{arXiv:1807.06211} [astro-ph.CO].
\bibitem{Hill-etal}
P. Ferreira, C. Hill and G. Ross, \PRD{95}{2017}{064038}
~\textsl{arXiv:1612.03157} [gr-qc].
\end{thebibliography}
\end{document}